\documentclass[a4paper,5p,fleqn]{cas-dc}
\usepackage[numbers]{natbib}
\usepackage{multirow}
\usepackage[table]{xcolor}
\usepackage{xcolor}
\definecolor{lightgreen}{RGB}{144, 238, 144}
\definecolor{lightorange}{RGB}{255, 200, 100}
\definecolor{lightred}{RGB}{255, 100, 100}
\usepackage{lineno}

\def\tsc#1{\csdef{#1}{\textsc{\lowercase{#1}}\xspace}}
\tsc{WGM}
\tsc{QE}
\tsc{EP}
\tsc{PMS}
\tsc{BEC}
\tsc{DE}

\begin{document}
\let\WriteBookmarks\relax
\def\floatpagepagefraction{1}
\def\textpagefraction{.001}
\shorttitle{Self-Supervised Ultrasound Representation Learning for Renal Anomaly Prediction in Prenatal Imaging}
\shortauthors{Y. Megahed \textit{et~al.}}

\title [mode = title]{Self-Supervised Ultrasound Representation Learning for Renal Anomaly Prediction in Prenatal Imaging}                      

\author[1,2]{Youssef Megahed}[type=editor,
                        orcid=0009-0004-2595-5468]
                        
\cormark[1]
\ead{youssefmegahed@cmail.carleton.ca}

\credit{Conceptualization of this study, Methodology, Software}

\affiliation[1]{organization={Department of Systems and Computer Engineering, Carleton University}, 
                city={Ottawa},
                state={Ontario},
                country={Canada}}
                
\affiliation[2]{organization={Department of Methodological and Implementation Research, Ottawa Hospital Research Institute}, 
                city={Ottawa},
                state={Ontario},
                country={Canada}}

\affiliation[3]{organization={Department of Acute Care Research, Ottawa Hospital Research Institute}, 
                city={Ottawa},
                state={Ontario},
                country={Canada}}

\affiliation[4]{organization={Children's Hospital of Eastern Ontario Research Institute}, 
                city={Ottawa},
                state={Ontario},
                country={Canada}}

\affiliation[5]{organization={Better Outcomes Registry \& Network Ontario, Children’s Hospital of Eastern}, 
                city={Ottawa},
                state={Ontario},
                country={Canada}}

\affiliation[6]{organization={Department of Obstetrics and Gynecology, University of Ottawa}, 
                city={Ottawa},
                state={Ontario},
                country={Canada}}

\affiliation[7]{organization={School of Epidemiology and Public Health, University of Ottawa}, 
                city={Ottawa},
                state={Ontario},
                country={Canada}}
                
\affiliation[8]{organization={Department of Obstetrics, Gynecology \& Newborn Care, The Ottawa Hospital}, 
                city={Ottawa},
                state={Ontario},
                country={Canada}}

\affiliation[9]{organization={International and Global Health Office, University of Ottawa}, 
                city={Ottawa},
                state={Ontario},
                country={Canada}}

\affiliation[10]{organization={Department of Clinical Science and Translational Medicine, University of Ottawa}, 
                city={Ottawa},
                state={Ontario},
                country={Canada}}

\author[3]{Inok Lee}
\author[3]{Robin Ducharme}
\author[4,5]{Kevin Dick}
\author[1]{Adrian D. C. Chan}
\author[1,2,4,7,10]{Steven Hawken}
\author[3,4,5,6,7,8,9]{Mark C. Walker}
\cormark[2]
\ead{mwalker@toh.ca}


\credit{Data curation, Writing - Original draft preparation}


\cortext[cor1]{Corresponding author}
\cortext[cor2]{Principal corresponding author}

\begin{abstract}
Prenatal ultrasound is the cornerstone for detecting congenital anomalies of the kidneys and urinary tract, but diagnosis is limited by operator dependence and suboptimal imaging conditions. We sought to assess the performance of a self-supervised ultrasound foundation model for automated fetal renal anomaly classification using a curated dataset of 969 two-dimensional ultrasound images. A pretrained Ultrasound Self-Supervised Foundation Model with Masked Autoencoding (USF-MAE) was fine-tuned for binary and multi-class classification of normal kidneys, urinary tract dilation, and multicystic dysplastic kidney. Models were compared with a DenseNet-169 convolutional baseline using cross-validation and an independent test set. USF-MAE consistently improved upon the baseline across all evaluation metrics in both binary and multi-class settings. USF-MAE achieved an improvement of about \textbf{1.87\%$\uparrow$} (AUC) and \textbf{7.8\%$\uparrow$} (F1-score) on the validation set, \textbf{2.32\%$\uparrow$} (AUC) and \textbf{4.33\%$\uparrow$} (F1-score) on the independent holdout test set. The largest gains were observed in the multi-class setting, where the improvement in AUC was \textbf{16.28\%$\uparrow$} and \textbf{46.15\%$\uparrow$} in F1-score. To facilitate model interpretability, Score-CAM visualizations were adapted for a transformer architecture and show that model predictions were informed by known, clinically relevant renal structures, including the renal pelvis in urinary tract dilation and cystic regions in multicystic dysplastic kidney. These results show that ultrasound-specific self-supervised learning can generate a useful representation as a foundation for downstream diagnostic tasks. The proposed framework offers a robust, interpretable approach to support the prenatal detection of renal anomalies and demonstrates the promise of foundation models in obstetric imaging.
\end{abstract}




\maketitle

\section{Introduction}
Clinical use of artificial intelligence (AI) has become deeply integrated into healthcare delivery and health system organization {\hypersetup{hidelinks}\textcolor{blue}{\cite{b1}}}. Deep learning (DL), which is a subset of machine learning (ML), has outperformed several other AI approaches, especially when it comes to data-intensive applications or identifying features that are not perceivable through human visual assessments {\hypersetup{hidelinks}\textcolor{blue}{\cite{b1,b2}}}. The DL of medical images has been one of the most widely developed applications, and convolutional neural networks (CNNs) are one of the most prevalent methods for medical image interpretation and classification {\hypersetup{hidelinks}\textcolor{blue}{\cite{b3,b4}}}.

One unique element of ultrasonography in comparison to other imaging modalities such as X-ray, CT, and MRI, is that image acquisition and interpretation are much more operator dependent, and therefore, experience a greater degree of image quality variability and limitations in generalizability {\hypersetup{hidelinks}\textcolor{blue}{\cite{b21,b22}}}. This creates unique challenges for DL model development and application {\hypersetup{hidelinks}\textcolor{blue}{\cite{b5}}}. Ultrasound is, however, particularly important in the field of obstetrics due to its unique ability to evaluate the pregnancy in real time and without radiation. The field of maternal-fetal medicine has therefore become a major center of focus for AI research {\hypersetup{hidelinks}\textcolor{blue}{\cite{b6}}}. While there has been a large amount of DL research on prenatal ultrasound examinations for standardized and protocolized tasks such as biometry and plane detection, there have been very few studies looking at structural anomalies, given that these are highly subjective and rely on expert-level interpretation {\hypersetup{hidelinks}\textcolor{blue}{\cite{b4,b7}}}.

Congenital anomalies of the kidneys and urinary tract (CAKUT) are among the most common fetal malformations, with an estimated prevalence of 1 in 500 live births, and account for 20-30\% of all prenatal anomalies {\hypersetup{hidelinks}\textcolor{blue}{\cite{b8,b9}}}. The kidney and urinary tract are intimately related to the production of amniotic fluid and lung development, and alterations in either can have significant clinical ramifications {\hypersetup{hidelinks}\textcolor{blue}{\cite{b10}}}. There are several CAKUT patterns that can be directly visualized during routine prenatal ultrasound examinations, but false positive findings are possible and can cause undue parental anxiety or lead to further investigations {\hypersetup{hidelinks}\textcolor{blue}{\cite{b12}}}. The diagnostic accuracy could therefore benefit from the integration of robust and reproducible image assessment.

\begin{figure*}
	\centering
	\includegraphics[width=.98\textwidth]{class_example_v2.jpeg}
	\caption{Prenatal ultrasound examples of renal development.
    (A) Normal fetal kidney with typical structure.
    (B) Multicystic dysplastic kidney (MCDK) showing multiple noncommunicating cystic spaces.
    (C) Urinary tract dilation (UTD) featuring enlarged renal pelvis and collecting system.}
	\label{FIG:class_example}
\end{figure*}

Our previous research developed and validated a DL model for detecting CAKUT in fetal ultrasound images {\hypersetup{hidelinks}\textcolor{blue}{\cite{b11}}}. That work introduced a convolutional classification framework trained on the same curated renal ultrasound dataset. It showed that a DenseNet-169 architecture could identify renal anomalies with good overall accuracy. It also demonstrated the importance of consistent preprocessing and annotation removal, and highlighted the challenges posed by class imbalance and subtle anatomical variation in prenatal imaging. Building on this work, we aim to extend more rigorous DL techniques to better detect fetal anomalies. Therefore, the objective of this study is to demonstrate the feasibility of a new DL model in detecting prenatal kidney anomalies. 

To advance beyond traditional supervised CNN-based pipelines used in our previous work, this study introduces a self-supervised ultrasound foundation model that learns modality-specific representations from large-scale, heterogeneous ultrasound datasets and then finetunes its Vision Transformer (ViT) encoder on the fetal renal ultrasound dataset. Recent progress in Self-Supervised Learning (SSL) has shown that models trained without manual labels can capture rich structural and textural features that transfer effectively to downstream clinical tasks {\hypersetup{hidelinks}\textcolor{blue}{\cite{b14,b17}}}. In this work, we leverage a pretrained Ultrasound Self-Supervised Foundation Model with Masked Autoencoding (USF-MAE), which was developed using more than 370,000 ultrasound images from the OpenUS-46 corpus ({\hypersetup{hidelinks}\textcolor{blue}{Fig.~\ref{FIG:model_arc}}}). This pretraining strategy allows the model to internalize diverse anatomical patterns, scanning conditions, and sonographic appearances, ultimately enhancing robustness in real prenatal imaging scenarios.

By fine-tuning the USF-MAE encoder for both binary and multi-class classification of fetal renal anomalies, we aim to evaluate whether this foundation-model approach can outperform a conventional convolutional architecture when trained on a smaller, curated dataset. Additionally, given the clinical importance of transparency in AI-assisted diagnosis, we integrate explainable AI (XAI) methods to interpret the decision-making process of the model. We adapted Score-CAM {\hypersetup{hidelinks}\textcolor{blue}{\cite{b20}}}, a class activation mapping technique, for transformer-based architectures to visualize the image regions that most contributed to the model's predictions. These visualization tools allow us to assess whether the model focuses on clinically relevant anatomical structures and to identify factors that may drive misclassification.

Through this combined use of self-supervised representation learning and XAI-guided assessment, the present study aims to demonstrate a novel and clinically meaningful framework for automated detection of fetal renal anomalies on prenatal ultrasound.

\section{Methodology}
\subsection{Study Setting and Data Acquisition}
This study used the same prenatal ultrasound dataset described in our original investigation {\hypersetup{hidelinks}\textcolor{blue}{\cite{b11}}}, although the present work applies different DL architectures. The dataset contains retrospectively acquired ultrasound images from singleton and twin pregnancies conducted at multiple sites of a tertiary care hospital in Ontario, Canada, between March 2014 and November 2021. This work was approved by the Ottawa Health Sciences Network Research Ethics Board (OHSN-REB \#20210079), and all procedures were performed in accordance with relevant institutional and regulatory guidelines.

Ultrasound images were collected by trained obstetric sonographers using GE Voluson V830 (916 images) and V730 (53 images) machines and interpreted by maternal-fetal medicine specialists. Normal images and images from cases with CAKUT taken between 18 and 24 weeks of gestation were selected, as fetal renal structures are identifiable by 18 weeks and pre-24-week diagnoses of urinary tract abnormalities can provide important prognostic information about postnatal renal function {\hypersetup{hidelinks}\textcolor{blue}{\cite{b12}}}.

\begin{table*}[htbp]
\centering
\caption{Data distribution across training, validation (4-fold cross-validation), and independent holdout test sets.}
\label{tab:data_split}
\setlength{\tabcolsep}{10pt}
\renewcommand{\arraystretch}{1.5}

\begin{tabular}{lcccc}
\toprule
 & \textbf{Overall, n (\%)} & \textbf{Normal, n (\%)} & \multicolumn{2}{c}{\textbf{Abnormal}} \\
\cmidrule(lr){4-5}
 &  &  & \textbf{MCDK, n (\%)} & \textbf{UTD, n (\%)} \\
\midrule

\textbf{Total}        & 969 (100\%)  & 646 (100\%)  & 64 (100\%)  & 259 (100\%) \\
\textbf{Training}     & 606 (62.54\%) & 404 (62.54\%) & 40 (62.50\%) & 162 (62.55\%) \\
\textbf{Validation}   & 202 (20.85\%) & 135 (20.90\%) & 13 (20.31\%) & 54 (20.85\%) \\
\textbf{Independent Holdout Test} & 161 (16.62\%) & 107 (16.56\%) & 11 (17.19\%) & 43 (16.60\%) \\
\bottomrule
\end{tabular}
\end{table*}

All images consisted of two-dimensional transverse abdominal planes, with measurement of the anteroposterior diameter of the renal pelvis. These images were retrieved from the institutional Picture Archiving and Communication System (PACS) and stored in Digital Imaging and Communications in Medicine (DICOM) format. Because the study focuses on a curated set of transverse fetal kidney images, the dataset was assembled to include normal kidneys and two CAKUT phenotypes that can be reliably identified in this specific plane: multicystic dysplastic kidney (MCDK) and urinary tract dilation (UTD) ({\hypersetup{hidelinks}\textcolor{blue}{Fig.~\ref{FIG:class_example}}} shows an example of each class). The combination of MCDK and UTD in the "Abnormal" meta-category is used to apply hierarchical categorization, converting the ML problem into a binary classification task ({\hypersetup{hidelinks}\textcolor{blue}{Fig.~\ref{FIG:class_types}}). UTD cases followed the 2014 classification criteria {\hypersetup{hidelinks}\textcolor{blue}{\cite{b12}}}, which set a renal pelvis diameter of 4 mm or greater as the diagnostic threshold. To ensure comparability with abnormal cases, normal samples were selected so that their acquisition years matched those of UTD and MCDK examinations. This produced a balanced and temporally aligned cohort of normal scans. The final normal class consisted of axial kidney images from pregnancies without CAKUT, resulting in an overall dataset ratio of about two normal cases for every abnormal case. Multiple examinations from the same patient were included when available within the gestational age window.

\begin{figure*}
	\centering
	\includegraphics[width=.7\textwidth]{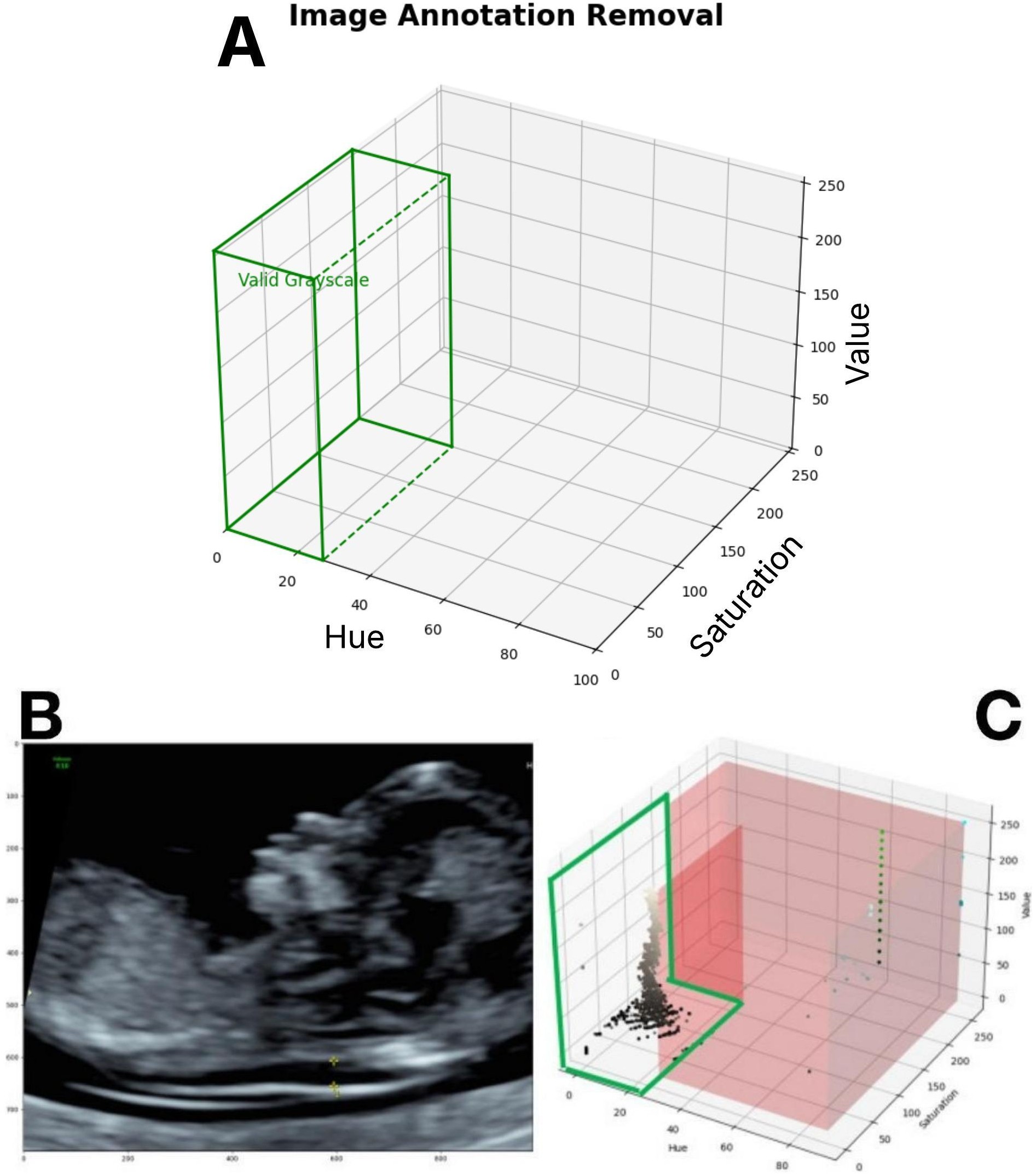}
	\caption{(A) Three-dimensional illustration of the allowable grayscale region in HSV space based on the thresholds for hue, saturation, and value.
    (B) Original ultrasound image containing colored annotations.
    (C) Visualization of the same image within HSV space, with pixels outside the grayscale region highlighted to show how the preprocessing method filters out annotation artifacts. Adapted from Walker et al.{\textcolor{blue}{\cite{b13}}} and {\textcolor{blue}{\cite{b11}}}.}
	\label{FIG:preprocessing_example}
\end{figure*}

\subsection{Data Preprocessing}
Our study adopted the same preprocessing pipeline applied in the earlier investigation that introduced this dataset, as shown in {\hypersetup{hidelinks}\textcolor{blue}{Fig.~\ref{FIG:preprocessing_example}}} {\hypersetup{hidelinks}\textcolor{blue}{\cite{b11}}}. The DICOM ultrasound images required varying levels of preparation before they could be used in our deep learning framework. Some images contained visual markup, including coloured callipers, measurement bars, text labels, icons, and profile traces ({\hypersetup{hidelinks}\textcolor{blue}{Fig.~\ref{FIG:preprocessing_before_after}A}}). Other images displayed patient identifiers or other forms of personal health information. Consistent with the previously published de-annotation and de-markup procedures {\hypersetup{hidelinks}\textcolor{blue}{\cite{b13}}}, we removed all coloured overlays and all visible personal information to avoid potential sources of bias or inadvertent leakage of class-specific indicators ({\hypersetup{hidelinks}\textcolor{blue}{Fig.~\ref{FIG:preprocessing_before_after}B}}). Each image was manually verified after these steps to ensure that only clean, unmarked content entered the modelling dataset. After annotation removal and deidentification, all images were added to the modelling dataset in three labelled categories: normal, UTD, and MCDK.

\begin{figure}
	\centering
	\includegraphics[width=.45\textwidth]{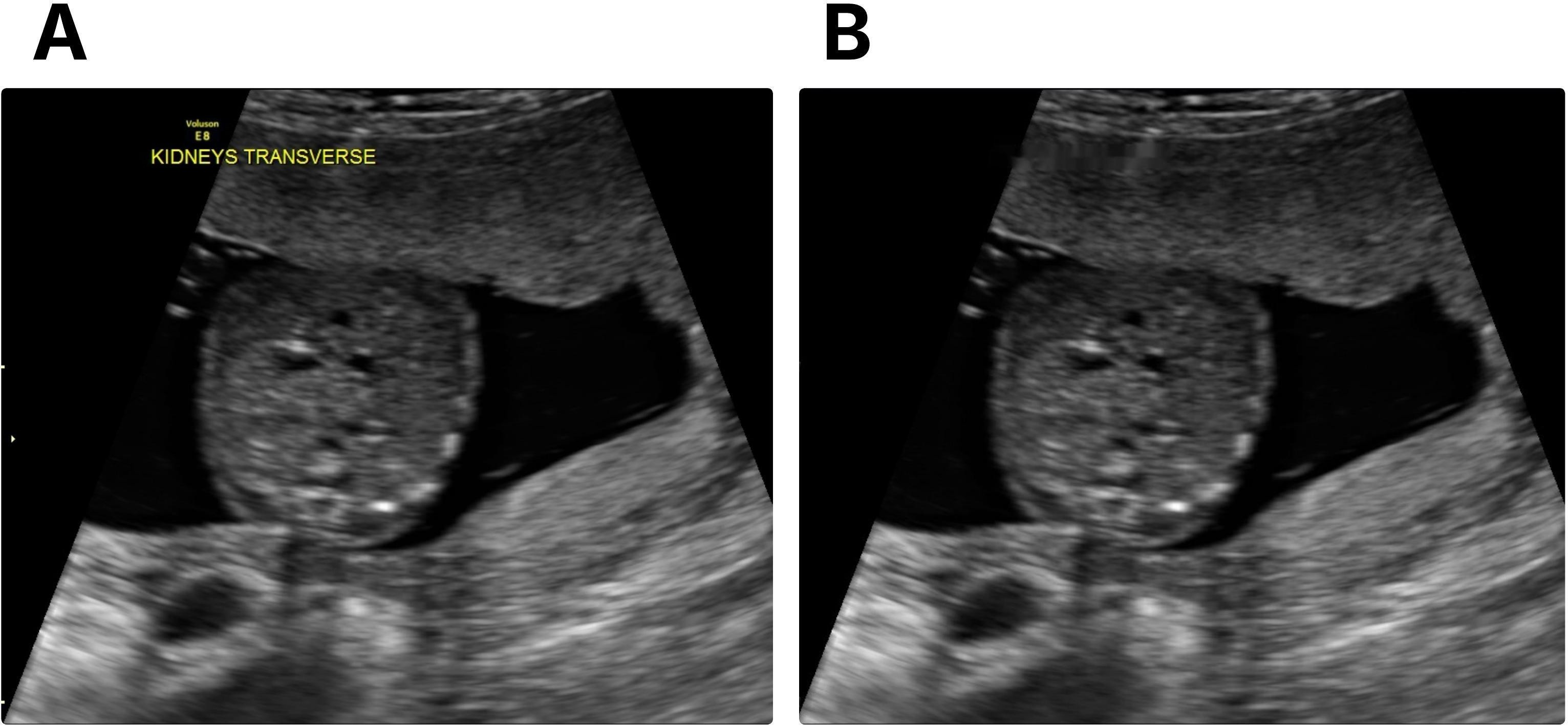}
	\caption{Example of annotation removal on a fetal renal ultrasound.
    (A) Original image containing overlaid text annotations.
    (B) Output after applying the preprocessing method to eliminate annotation artifacts while preserving underlying anatomical details.}
	\label{FIG:preprocessing_before_after}
\end{figure}

\subsection{Dataset Distribution}
The final set contained a total of 969 fetal kidney ultrasound images, of which 646 were normal, and 323 were abnormal (64 MCDK and 259 UTD), as summarized in {\hypersetup{hidelinks}\textcolor{blue}{Table~\ref{tab:data_split}}}. Stratified sampling was first applied to create two disjoint subsets: an independent holdout test set and a cross-validation set ({\hypersetup{hidelinks}\textcolor{blue}{Table~\ref{tab:data_split}}}). The independent test set contained 161 images ($\backsim$17\% of the whole dataset) and was reserved solely for final model evaluation. The remaining 808 images formed the cross-validation set. This set was later partitioned into four stratified folds for model development. In each cross-validation cycle, three folds were used for training and one fold for validation, with all folds serving as the validation fold once.

Stratification was performed at the class level to ensure the same proportional split of normal and abnormal cases across each subset. As such, about 62.5\% of each class was used for training folds, 20\% for the validation fold, and the remaining ~17.5\% for the final test set. This sampling technique provided an equal representation of majority and minority classes at all stages of model development.

\subsection{Model Training Framework}
\textbf{1) USF-MAE Architecture:}
USF-MAE {\hypersetup{hidelinks}\textcolor{blue}{\cite{b14}}} is a foundation transformer-based SSL model that was developed by our research group, designed to learn high-level and modality-specific features directly from unlabeled ultrasound images {\hypersetup{hidelinks}\textcolor{blue}{\cite{b14,b17}}}. It follows the Masked Autoencoder framework introduced by He \textit{et al.} {\hypersetup{hidelinks}\textcolor{blue}{\cite{b15}}} but is adapted for ultrasound imaging. An overview of this architecture is shown in {\hypersetup{hidelinks}\textcolor{blue}{Fig.~\ref{FIG:model_arc}}}.

During pretraining, each ultrasound image is partitioned into fixed, non-overlapping $16 \times 16$ patches. A subset of these patches, corresponding to $25\%$ of all patches, is hidden at random. The visible patches are then linearly projected to embedding vectors, which are fed into the transformer-based (ViT) encoder. The encoder, trained with multi-head self-attention, learns local spatial dependencies and texture cues distinctive to ultrasound imaging data {\hypersetup{hidelinks}\textcolor{blue}{\cite{b14,b17}}}.

The decoder receives both the encoded visible patches and learned tokens marking the masked positions. It reconstructs the original image from contextual cues. Pretraining is optimized by minimizing the Mean Squared Error (MSE) between the reconstructed image $\hat{x}$ and the original image $x$:

\begin{equation}
\mathcal{L}_{\text{MAE}} = \frac{1}{N} \sum_{i=1}^{N} \left\| \hat{x}_i - x_i \right\|_2^2,
\end{equation}

\noindent
where $N$ denotes the total number of pixels in the image. The OpenUS-46 corpus used for pretraining contains approximately 370,000 ultrasound scans from 46 publicly available datasets {\hypersetup{hidelinks}\textcolor{blue}{\cite{b14}}}, providing a diverse source of anatomical variation that serves as a suitable foundation for downstream fetal renal classification.

The renal anomaly classification models were developed by fine-tuning the USF-MAE encoder {\hypersetup{hidelinks}\textcolor{blue}{\cite{b14}}} on the curated fetal kidney ultrasound dataset, following the same general training strategy outlined in our recent work in {\hypersetup{hidelinks}\textcolor{blue}{\cite{b17}}}. The pretrained encoder served as the initialization point for both the binary task (normal versus abnormal) and the multi-class task (normal, MCDK, and UTD). This class structure is illustrated in {\hypersetup{hidelinks}\textcolor{blue}{Fig.~\ref{FIG:class_types}}}. Model initialization, supervised fine-tuning with cross-validation, hyperparameter tuning, and evaluation on the hold-out test set comprise the training workflow. The computational environment and software stack were kept consistent for all experiments to ensure reproducibility.

\textbf{2) Initialization of the Model:}
All experiments began with the pretrained USF-MAE encoder obtained from the OpenUS-46 self-supervised training phase. Since the reconstruction decoder is only required during pretraining, it was removed during fine-tuning. For the binary renal anomaly task, a fully-connected layer was added to map the encoder representation to two output units (normal and abnormal). For the multi-class task, the classification head contained three output units corresponding to normal, MCDK, and UTD. A SoftMax function was applied to convert logits into normalized class probabilities. The encoder and classification head were fine-tuned jointly.

\begin{figure}
	\centering
	\includegraphics[width=.45\textwidth]{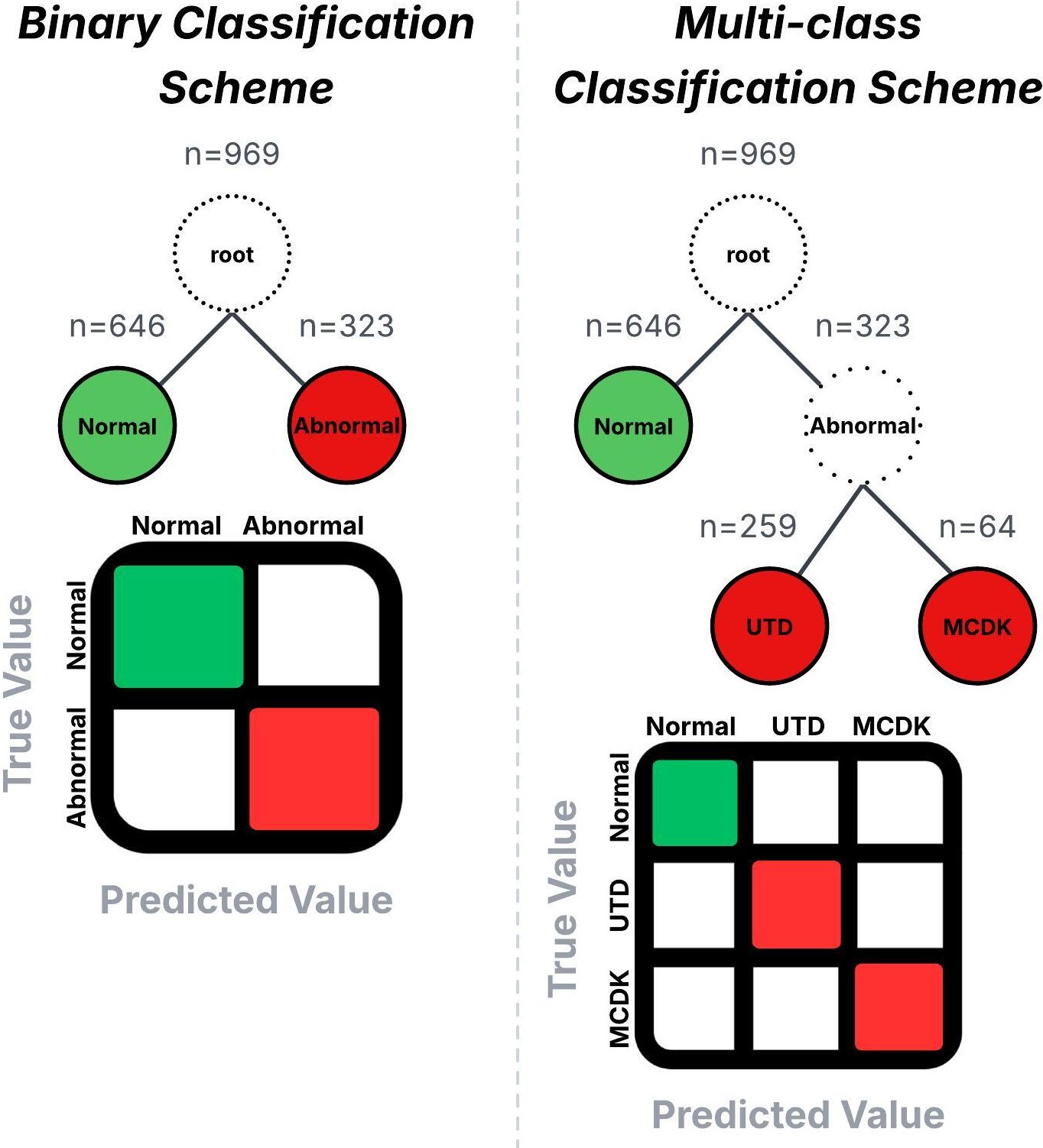}
	\caption{Comparison of the binary and multi-class classification schemes.
    The left panel shows the binary scheme, where images are hierarchically organized into normal or abnormal classes, with a corresponding two-class confusion matrix. The right panel illustrates the multi-class approach, in which abnormal cases are further separated into UTD and MCDK, resulting in a three-class prediction structure and its associated confusion matrix.}
	\label{FIG:class_types}
\end{figure}

\textbf{3) Supervised Fine-tuning Framework:}
The dataset was divided into training, validation, and independent holdout test sets using the stratified procedure described in Section~2.3. A test set was reserved in advance of model training and used only once at the end to assess model performance. The remaining dataset was then split into folds for cross-validation. Care was taken to ensure that the class balance was identical in all folds.

For each experiment, a 4-fold cross-validation strategy was applied. Three of these folds were used as the training set, and one fold as the validation set. The 4-fold CV procedure was repeated 5 times with a different stratified split of images in each repetition, to give a more robust estimate of model performance. Results from all repetitions were then averaged to produce the final validation metrics. For both binary and multi-class tasks, class weighting was applied to the cross-entropy loss to address imbalance among normal, MCDK, and UTD cases. Class weights were computed as the inverse of the class frequencies observed within each training fold.

\begin{figure*}
\centering
\includegraphics[width=.91\textwidth]{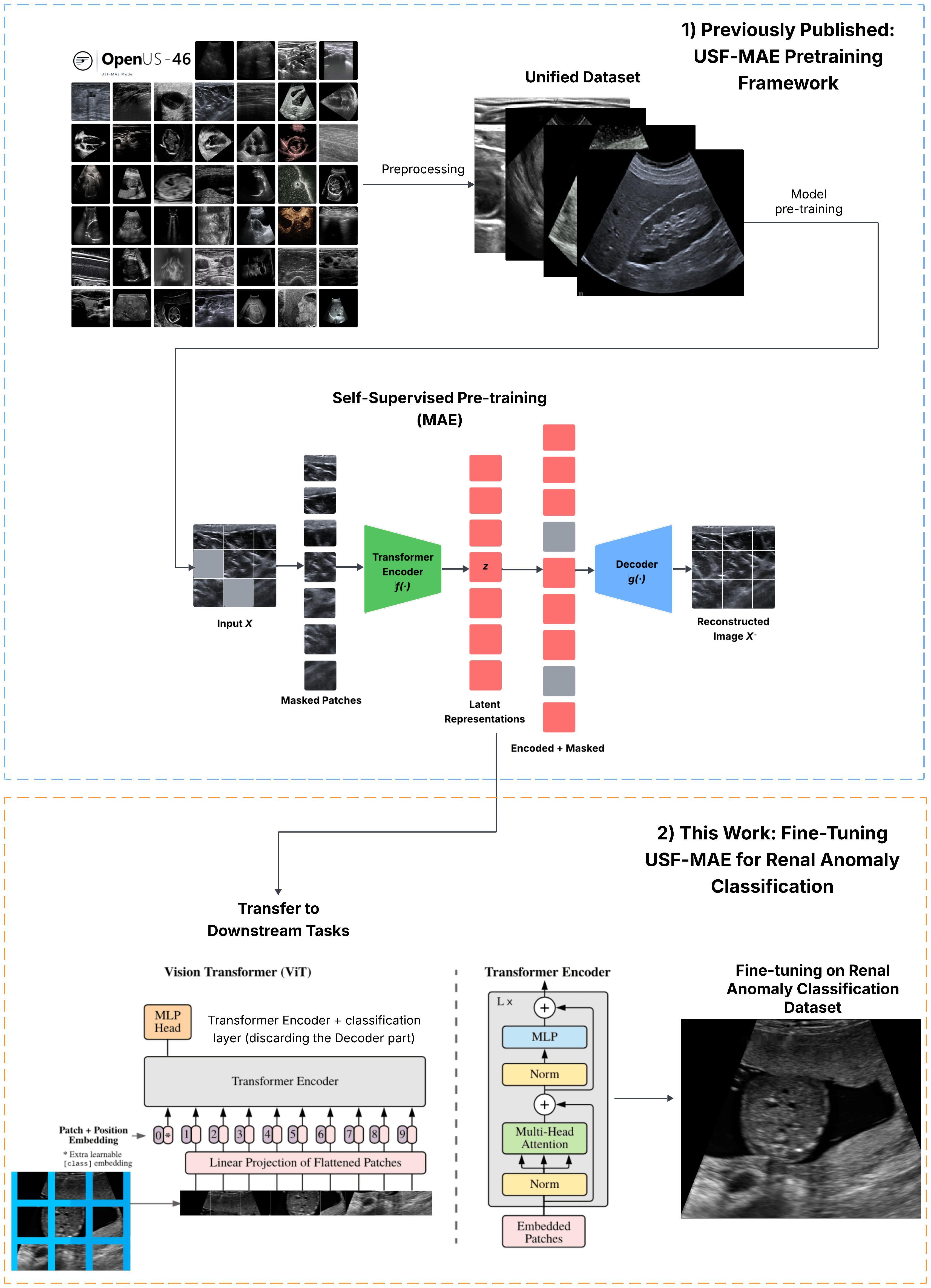}
\caption{Summary of the renal anomaly classification workflow. The top panel outlines the previously developed USF-MAE pretraining framework {\textcolor{blue}{\cite{b14}}}, in which a large, unified ultrasound dataset is preprocessed and used for SSL with a masked autoencoder. Random image patches are hidden and reconstructed during training, allowing the encoder to capture ultrasound-specific feature representations. The bottom panel shows this study's extension, in which the pretrained encoder is adapted for renal anomaly classification by removing the MAE decoder and adding a classification head. The resulting model is fine-tuned with labelled renal ultrasound data to support diagnostic prediction.}
\label{FIG:model_arc}
\end{figure*}

\textbf{4) Optimization Strategy and Hyperparameter Search: } Fine-tuning was performed using the AdamW optimizer {\hypersetup{hidelinks}\textcolor{blue}{\cite{b16}}} to implement an adaptive step size while also incorporating weight decay regularization to stabilize the optimization process for transformer-based architectures. Grid search was conducted for hyperparameter tuning, using a grid of learning rates $\{0.001, 0.0003, 0.0005, 0.00001\}$ and weight decay values $\{0.01, 0.05, 0.001, 0.0001\}$. 4-fold cross-validation was used to train and validate each parameter combination, and the best-performing combination was selected as a learning rate of $0.0003$, weight decay of $0.01$, and a batch size of $64$ with $224 \times 224$ input image size for both binary and multi-class models.

A cosine learning rate schedule with linear warm-up over the first $10\%$ of epochs was applied, followed by smooth decay for the remainder of training. Gradient clipping with a maximum norm of $1.0$ was applied to the model weights to avoid unstable updates.

During cross-validation, the model state with the highest validation accuracy was saved and used as a fold-specific checkpoint. The metrics from each fold-specific checkpoint were averaged to obtain the reported final results.

\textbf{5) Computational Environment and Software Stack:}
The model was implemented using PyTorch, and the USF-MAE encoder was loaded using the Hugging Face Transformers API. NumPy, Pandas and scikit-learn are used for data manipulation and evaluation.

Training was performed on a SLURM-managed high-performance research computing cluster with NVIDIA L40S GPUs; a single fold required 26 minutes of training time over 100 epochs. Checkpoints, logs, and configuration files were all stored to enable full reproducibility. The pretrained USF-MAE weights and the OpenUS-46 ultrasound image corpus have been made publicly available in our repository (\href{https://github.com/Yusufii9/USF-MAE}{\textcolor{blue}{GitHub Repository}}: https://github.com/Yusufii9/USF-MAE).

\begin{table*}[htbp]
\centering
\caption{Comparative performance of baseline and USF-MAE model on renal anomaly \textbf{binary classification (validation performance)}.}
\label{tab:val_set_binary}
\setlength{\tabcolsep}{2.5pt}
\renewcommand{\arraystretch}{1.5}

\begin{tabular}{lcccccc}
\toprule
\rowcolor[gray]{0.85}
\textbf{Model} &
\textbf{AUC} &
\textbf{Accuracy} &
\textbf{F1-score} &
\textbf{Recall} &
\textbf{Specificity} &
\textbf{Precision} \\

\hline

DenseNet-169 {\hypersetup{hidelinks}\textcolor{blue}{\cite{b11}}} &
0.9128 $\pm$ 0.0052 &
0.8403 $\pm$ 0.0076 &
0.7630 $\pm$ 0.017 &
0.7739 $\pm$ 0.0199 &
0.8735 $\pm$ 0.0128 &
0.7580 $\pm$ 0.0231 \\

\textbf{USF-MAE (new)} &
\textbf{0.9315 $\pm$ 0.0211} &
\textbf{0.8988 $\pm$ 0.0186} &
\textbf{0.8410 $\pm$ 0.0324} &
\textbf{0.8074 $\pm$ 0.0513} &
\textbf{0.9443 $\pm$ 0.0208} &
\textbf{0.8807 $\pm$ 0.0387} \\
\bottomrule
\end{tabular}
\end{table*}

\subsection{Baseline Model: DenseNet-169} 
In order to give a convolutional baseline for comparison, our USF-MAE framework was compared to the DenseNet-169 architecture from {\hypersetup{hidelinks}\textcolor{blue}{\cite{b11}}}, similar to the previous work on cystic hygroma classification {\hypersetup{hidelinks}\textcolor{blue}{\cite{b18}}}. DenseNet is a family of convolutional neural networks {\hypersetup{hidelinks}\textcolor{blue}{\cite{b19}}} in which each layer is directly connected to every other layer in a feed-forward fashion, enabling the network to reuse features and maintain low computational cost while maintaining strong gradient flow, which is necessary for ultrasound image processing. In the work mentioned above, a PyTorch implementation of DenseNet-169 was used and the first convolutional layer was adapted to work with 1-channel grayscale image inputs at different spatial resolutions. The final classifier was also replaced with a fully connected layer with 2 or 3 output nodes, depending on the binary or multiclass classification task.

The network was trained by minimizing a weighted cross-entropy loss function. The weights were set to the inverse of class counts to handle class imbalance and were further normalized by the specific binary or three-class classification setup.

To find the optimal DenseNet-169 configuration, an extensive search was conducted over a set of hyperparameters. The authors considered three image input resolutions: $128 \times 128$, $256 \times 256$, and $512 \times 512$. This allowed the determination of the effect of spatial resolution on final accuracy while trading off computational costs. The authors also investigated three different numbers of epochs to train for: 100, 300, and 600. This determined whether prolonged training was needed or if a saturation point was quickly reached. Furthermore, three different batch sizes were tested: 32, 64, and 128, in order to see the effect of training stability and resource utilization on overall training.

The network was trained on all hyperparameter combinations using a step-decay learning rate schedule. A starting learning rate of 0.1 was used, with a step size of 55 epochs and a decay constant of $\gamma = 0.3$, resulting in a smooth learning rate decay.

Out of all the hyperparameter combinations, the authors found that the DenseNet-169 model with an image size of $128 \times 128$ pixels, 300 training epochs, and a batch size of 32 achieved the best performance and was therefore used as the final DenseNet-169 model for comparison with our new proposed framework.

\subsection{Performance Metrics}
To evaluate the quantitative model performance for both the proposed framework (USF-MAE) as well as baseline architecture (DenseNet-169), we assessed the following classification metrics: accuracy, precision, recall, F1-score, specificity, and AUC. Collectively, these describe performance in terms of classification accuracy, diagnostic sensitivity, and normal vs. abnormal renal ultrasound scan detection. As the clinical use-case for this work is to provide accurate detection of fetal renal anomalies to support robust and reliable identification of affected fetuses, relevant priority is given to the model's performance in terms of detecting abnormal cases and minimizing false-negative results. All metrics were computed on the validation set and heldout test set, with final reported results corresponding to mean performance across cross-validation folds (5 repeats).

\textbf{1) Accuracy: } Accuracy is a standard metric that measures the overall proportion of correctly classified samples across both classes. 
\begin{equation}
\text{Accuracy} = \frac{TP + TN}{TP + TN + FP + FN}
\end{equation} 
TP, TN, FP, and FN represent true positives, true negatives, false positives, and false negatives, respectively. Note that while accuracy represents a simple way to assess overall model performance, it is not sufficient on its own when the dataset is imbalanced, which is often the case in many clinical applications.

\textbf{2) Precision: } Precision is also known as Positive Predictive Value (PPV) and describes the fraction of predicted abnormal cases that are truly abnormal. \begin{equation}
\text{Precision} = \frac{TP}{TP + FP}
\end{equation}
In the clinical context of this work, a high-precision model is desired in order to minimize false-positive predictions. False positives would lead to an increased number of follow-up imaging referrals or unnecessary specialist clinic attendances.

\begin{table*}[htbp]
\centering
\caption{Comparative performance of baseline and USF-MAE model on renal anomaly \textbf{binary classification (independent holdout test performance)}.}
\label{tab:test_set_binary}
\setlength{\tabcolsep}{2.5pt}
\renewcommand{\arraystretch}{1.5}

\begin{tabular}{lcccccc}
\toprule
\rowcolor[gray]{0.85}
\textbf{Model} &
\textbf{AUC} &
\textbf{Accuracy} &
\textbf{F1-score} &
\textbf{Recall} &
\textbf{Specificity} &
\textbf{Precision} \\

\hline

DenseNet-169 {\hypersetup{hidelinks}\textcolor{blue}{\cite{b11}}} &
0.9071 $\pm$ 0.0054 &
0.8170 $\pm$ 0.0088 &
0.7488 $\pm$ 0.0187 &
\textbf{0.8120 $\pm$ 0.0240} &
0.8206 $\pm$ 0.0174 &
0.7007 $\pm$ 0.0231 \\

\textbf{USF-MAE (new)} &
\textbf{0.9303 $\pm$ 0.0131} &
\textbf{0.8646 $\pm$ 0.0209} &
\textbf{0.7921 $\pm$ 0.0344} &
0.7722 $\pm$ 0.0620 &
\textbf{0.9112 $\pm$ 0.0304} &
\textbf{0.8183 $\pm$ 0.0473} \\
\bottomrule
\end{tabular}
\end{table*}

\textbf{3) Recall (Sensitivity): } Recall is the proportion of true abnormal cases that were identified by the model. 
\begin{equation}
\text{Recall} = \frac{TP}{TP + FN}
\end{equation} 
In the screening setting of this work, a high-recall model is critical in order to minimize the false-negative rate. A false negative would lead to a missed opportunity to follow up on a renal anomaly in a timely manner and plan further investigation and early management.

\textbf{4) F1-Score: } The F1-score is the harmonic mean of precision and recall. 
\begin{equation}
\text{F1-score} = \frac{2 \times (\text{Precision} \times \text{Recall})}{\text{Precision} + \text{Recall}}
\end{equation} 
As the F1-score penalizes models when both precision and recall are low, it is a strong overall performance metric to examine when the class distributions are not equal. A high F1-score therefore indicates that a model is detecting the majority of abnormal cases and is also making a limited number of false-positive predictions.

\textbf{5) Specificity: } Specificity measures the model’s ability to classify normal renal ultrasound images correctly. 
\begin{equation}
\text{Specificity} = \frac{TN}{TN + FP}
\end{equation} 
In the screening context, a high-specificity model is needed in order to correctly identify the majority of normal fetuses and not cause undue clinical concern or subsequent follow-up.

\textbf{6) Area Under the ROC Curve (ROC-AUC): } Area Under the Receiver Operating Characteristic (ROC) Curve (ROC-AUC) is a threshold-independent way of evaluating model separability. The ROC curve is a graphical plot of the true positive rate (recall) vs. the false positive rate (1-specificity) at all decision thresholds. The area under the ROC curve (AUC) ranges from 0 to 1, with 0.5 being equivalent to chance-level performance and 1 corresponding to a perfect classification model. High AUC values reflect the model's ability to consistently assign a higher confidence score to true abnormal cases vs. normal cases.

\begin{figure*}
	\centering
	\includegraphics[width=.98\textwidth]{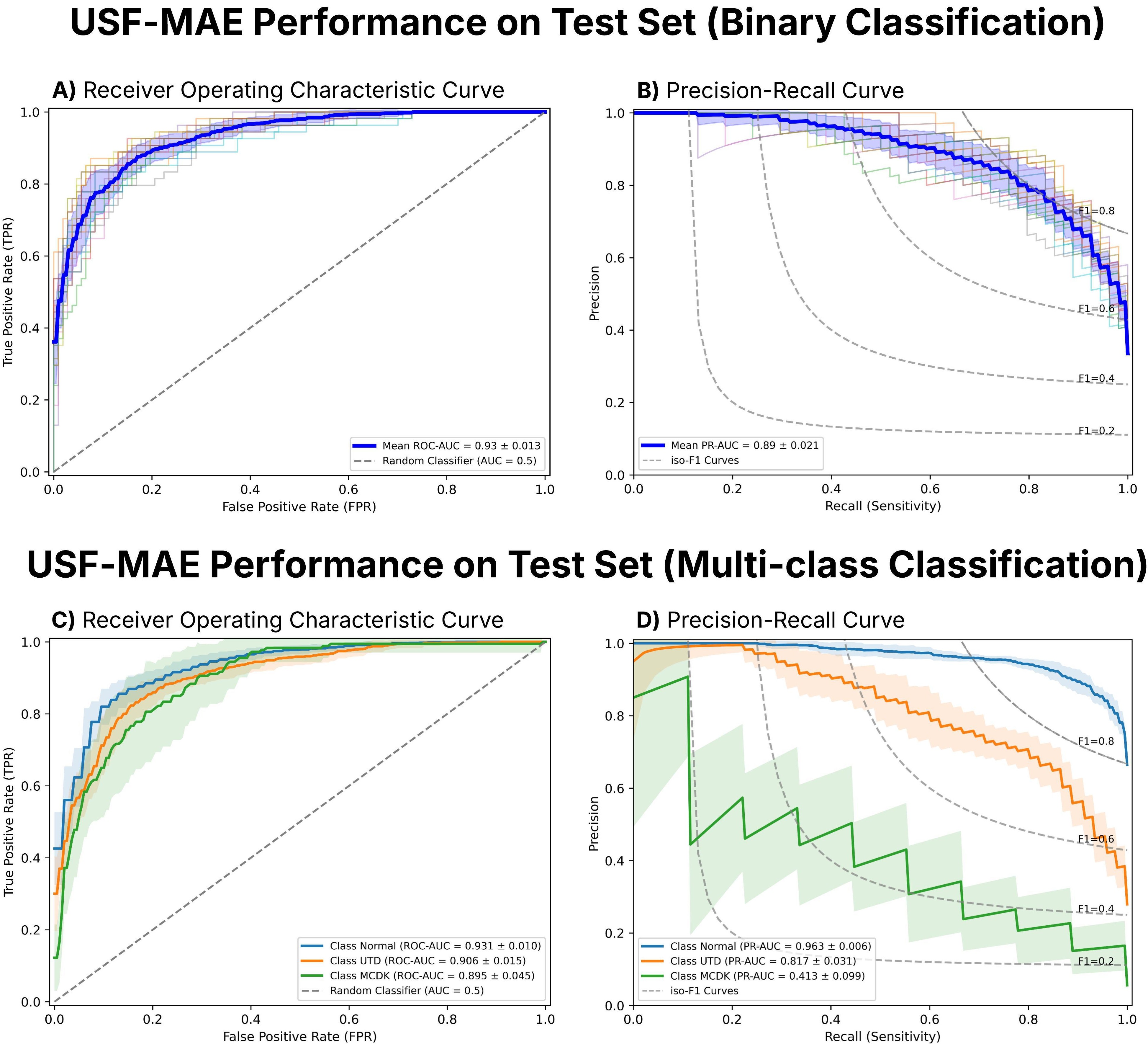}
	\caption{Test set performance of the USF MAE model for binary and multi-class renal anomaly classification.
    (A) ROC curve for the binary task, showing strong separation between positive and negative classes.
    (B) Precision-recall curve for the binary task, demonstrating high precision across a broad sensitivity range.
    (C) ROC curves for the multi-class setting, with separate performance shown for normal, UTD, and MCDK classes.
    (D) Precision-recall curves for the multi-class setting, highlighting variability in class-specific performance and illustrating the effect of class imbalance.}
	\label{FIG:performance_curves}
\end{figure*}

\subsubsection*{Multi-class Performance Metrics}
For the multi-class renal anomaly classification task, performance was evaluated using weighted precision, recall, and F1 score. Let $K$ denote the total number of classes, and for each class $k \in \{1, \dots, K\}$, let $TP_k$, $FP_k$, and $FN_k$ represent the true positives, false positives, and false negatives for that class. The support of each class, $n_k$, corresponds to the number of true samples belonging to class $k$, with weights defined as
\begin{equation}
w_k = \frac{n_k}{\sum_{i=1}^{K} n_i}.
\end{equation}

\textbf{Weighted Precision:} The weighted precision is computed as the weighted sum of class specific precision values
\begin{equation}
\text{Precision}_{k} = \frac{TP_k}{TP_k + FP_k},
\end{equation}
\begin{equation}
\text{Precision}_{weighted} = \sum_{k=1}^{K} w_k \cdot \text{Precision}_{k}.
\end{equation}

\textbf{Weighted Recall:} Similarly, weighted recall aggregates class specific recall values
\begin{equation}
\text{Recall}_{k} = \frac{TP_k}{TP_k + FN_k},
\end{equation}
\begin{equation}
\text{Recall}_{weighted} = \sum_{k=1}^{K} w_k \cdot \text{Recall}_{k}.
\end{equation}

\textbf{Weighted F1-Score:} The F1-score for each class is defined as
\begin{equation}
F1_{k} = \frac{2 \cdot \text{Precision}_{k} \cdot \text{Recall}_{k}}{\text{Precision}_{k} + \text{Recall}_{k}},
\end{equation}
and the weighted F1-score is given by
\begin{equation}
F1_{weighted} = \sum_{k=1}^{K} w_k \cdot F1_{k}.
\end{equation}

These weighted metrics account for class imbalance by incorporating the relative frequency of each class, ensuring that performance reflects clinically meaningful differences in prediction accuracy across Normal, UTD, and MCDK classes.

\subsection{Explainable AI Framework}
To visualize the regions most influential to model predictions, we incorporated an XAI analysis using Score-CAM {\hypersetup{hidelinks}\textcolor{blue}{\cite{b20}}}. Since the USF-MAE encoder is based on a ViT, which does not contain convolutional layers, Score-CAM was applied using a ViT-adapted procedure.

For each input image, the hidden token embeddings from the final encoder block were extracted from the output of the \texttt{layernorm\_before} operation. The class token was removed, and the remaining patch tokens were reshaped into a two-dimensional feature grid matching the spatial patch arrangement of the input image. This grid served as the activation map input for Score-CAM. A reshape transform was applied to reorder the token embeddings into the format required by the CAM framework.

Score-CAM attribution maps were then generated by creating masked versions of the input image for each feature map, passing these masked images through the classification head, and using the corresponding class scores as weights. The weighted combination of feature maps produced a high-resolution saliency map, which was subsequently upsampled and superimposed onto the original ultrasound image.

This XAI process was applied to correctly and incorrectly classified images in both binary and multi-class tasks. The resulting Score-CAM visualizations consistently highlighted clinically meaningful structures, including the renal pelvis in UTD cases and cystic regions in MCDK, demonstrating that model predictions were driven by anatomically relevant features rather than spurious artifacts.

\section{Results}
\subsection{Binary Classification Performance}
{\hypersetup{hidelinks}\textcolor{blue}{Tables~\ref{tab:val_set_binary}}} and {\hypersetup{hidelinks}\textcolor{blue}{\ref{tab:test_set_binary}}} report comparative results on the DenseNet-169 baseline model and the proposed USF-MAE model for binary classification between normal and abnormal fetal renal ultrasound volumes. For both the validation set and held-out test set, USF-MAE achieved higher performance on almost every evaluation metric than DenseNet-169.

In the validation set results ({\hypersetup{hidelinks}\textcolor{blue}{Table~\ref{tab:val_set_binary}}}), USF-MAE produced the best AUC of $0.9315 \pm 0.0211$ while DenseNet-169 achieved an AUC of $0.9128 \pm 0.0052$. USF-MAE also led to higher accuracy of $0.8988 \pm 0.0186$ and F1-score of $0.8410 \pm 0.0324$. In terms of class separation, USF-MAE achieved a recall of $0.8074 \pm 0.0513$ and a specificity of $0.9443 \pm 0.0208$, indicating the model's ability to detect abnormal renal cases with limited false positives. The USF-MAE model also led to improved precision of $0.8807 \pm 0.0387$ over the DenseNet-169 baseline.

USF-MAE's performance advantage over the DenseNet-169 baseline model was confirmed by results on the independent holdout test set ({\hypersetup{hidelinks}\textcolor{blue}{Table~\ref{tab:test_set_binary}}}). USF-MAE achieved higher performance than DenseNet-169 on almost every evaluation metric including AUC, where it achieved an AUC of $0.9303 \pm 0.0131$ compared to $0.9071 \pm 0.0054$. USF-MAE also achieved higher accuracy of $0.8646 \pm 0.0209$ and F1-score of $0.7921 \pm 0.0344$ compared to the DenseNet-169 baseline. In terms of class separation, USF-MAE achieved a recall of $0.7722 \pm 0.0620$, which is lower than DenseNet-169 by around 4\%, and a specificity of $0.9112 \pm 0.0304$, indicating better ability to detect abnormal renal cases and limiting false positives compared to DenseNet-169. The USF-MAE model also had higher precision of $0.8183 \pm 0.0473$ compared to the DenseNet-169 baseline ($0.7007 \pm 0.0231$).

\begin{table*}[htbp]
\centering
\caption{Comparative performance of DenseNet-169 and USF-MAE on \textbf{multi-class renal anomaly classification for the validation and independent test sets}.}
\label{tab:multi_class_combined}
\setlength{\tabcolsep}{2.5pt}
\renewcommand{\arraystretch}{1.5}

\begin{tabular}{lcccccc}
\toprule
\rowcolor[gray]{0.85}
\textbf{Validation Set} &
\textbf{AUC} &
\textbf{Accuracy} &
\textbf{F1-score} &
\textbf{Recall} &
\textbf{Specificity} &
\textbf{Precision} \\
\midrule

DenseNet-169 {\hypersetup{hidelinks}\textcolor{blue}{\cite{b11}}} &
0.7546 $\pm$ 0.0151 &
0.7246 $\pm$ 0.0118 &
0.3820 $\pm$ 0.0436 &
0.4148 $\pm$ 0.0250 &
0.8605 $\pm$ 0.0153 &
0.3694 $\pm$ 0.0469 \\

\textbf{USF-MAE (new)} &
\textbf{0.9174 $\pm$ 0.0191} &
\textbf{0.8488 $\pm$ 0.0170} &
\textbf{0.8435 $\pm$ 0.0194} &
\textbf{0.8488 $\pm$ 0.0170} &
\textbf{0.9003 $\pm$ 0.0158} &
\textbf{0.8465 $\pm$ 0.0195} \\

\midrule
\rowcolor[gray]{0.85}
\textbf{Test Set} &
\textbf{AUC} &
\textbf{Accuracy} &
\textbf{F1-score} &
\textbf{Recall} &
\textbf{Specificity} &
\textbf{Precision} \\
\midrule

\textbf{USF-MAE (new)} &
\textbf{0.9218 $\pm$ 0.0101} &
\textbf{0.8205 $\pm$ 0.0275} &
\textbf{0.8197 $\pm$ 0.0280} &
\textbf{0.8205 $\pm$ 0.0275} &
\textbf{0.8836 $\pm$ 0.0228} &
\textbf{0.8223 $\pm$ 0.0288} \\

\bottomrule
\end{tabular}
\end{table*}

\subsection{Multi-Class Classification Performance}
{\hypersetup{hidelinks}\textcolor{blue}{Table~\ref{tab:multi_class_combined}}} reports the results for the three-class classification of renal anomalies where the model needed to differentiate among normal kidneys, MCDK, and UTD subtypes. The performance of USF-MAE and DenseNet-169 is comparable to that observed for binary classification, with USF-MAE outperforming DenseNet-169 across all evaluation metrics.

USF-MAE's AUC was $0.9174 \pm 0.0191$, substantially higher than the DenseNet-169 baseline, which was $0.7546 \pm 0.0151$. USF-MAE's accuracy was $0.8488 \pm 0.0170$, which was higher than the DenseNet-169 baseline of $0.7246 \pm 0.0118$. USF-MAE's F1-score was $0.8435 \pm 0.0194$, which is considerably higher than the DenseNet-169 baseline of $0.3820 \pm 0.0436$. USF-MAE also achieved a recall of $0.8488 \pm 0.0170$ and a precision of $0.8465 \pm 0.0195$, both higher than the DenseNet-169 baseline. Lastly, USF-MAE achieved a specificity of $0.9003 \pm 0.0158$, which is higher than the DenseNet-169 baseline.

Performance on the independent holdout test set ({\hypersetup{hidelinks}\textcolor{blue}{Table~\ref{tab:multi_class_combined}}}) followed a similar trend. USF-MAE achieved an AUC of $0.9218 \pm 0.0101$, accuracy of $0.8205 \pm 0.0275$, F1-score of $0.8197 \pm 0.0280$, recall of $0.8205 \pm 0.0275$, specificity of $0.8836 \pm 0.0228$, and precision of $0.8223 \pm 0.0288$, indicating consistent generalization to unseen data. The baseline DenseNet-169 model from {\hypersetup{hidelinks}\textcolor{blue}{\cite{b11}}} did not report multi-class performance on the independent holdout test set, but given its substantially lower validation performance, its test-set results would be expected to be similarly limited or lower.

In summary, for the task of three-class renal anomaly classification, the USF-MAE model outperformed the baseline model (DenseNet-169), demonstrating superior performance across all evaluation metrics. Compared to the binary classification setting, the AUC, accuracy, and F1-score for the DenseNet-169 baseline model decreased significantly for the more challenging multi-class setting. However, with USF-MAE, the performance of the model on all metrics remained high despite the increase in task difficulty.

\subsection{Summary of Model Performance}
Across both the binary and multi-class fetal renal anomaly classification tasks, the proposed USF-MAE model outperformed the DenseNet-169 baseline model, obtaining higher AUC, accuracy, F1-score, precision, and specificity values. USF-MAE demonstrates a performance gain of \textbf{1.87\%$\uparrow$} (AUC) and \textbf{7.8\%$\uparrow$} (F1-score) on the validation set, \textbf{2.32\%$\uparrow$} (AUC) and \textbf{4.33\%$\uparrow$} (F1-score) on the holdout test set, and \textbf{16.28\%$\uparrow$} (AUC) and \textbf{46.15\%$\uparrow$} (F1-score) in the multi-class setting compared to DenseNet-169. These performance increases are especially notable for anomaly detection tasks such as fetal renal anomaly classification, where USF-MAE's advantage allows better discrimination of subtle anatomical differences between normal and abnormal cases. The generalization of this performance gain across both validation and held-out test sets highlights the robustness of USF-MAE self-supervised pretraining on ultrasound images and then fine-tuned on fetal renal anomaly classification. 

\begin{figure*}
	\centering
	\includegraphics[width=.98\textwidth]{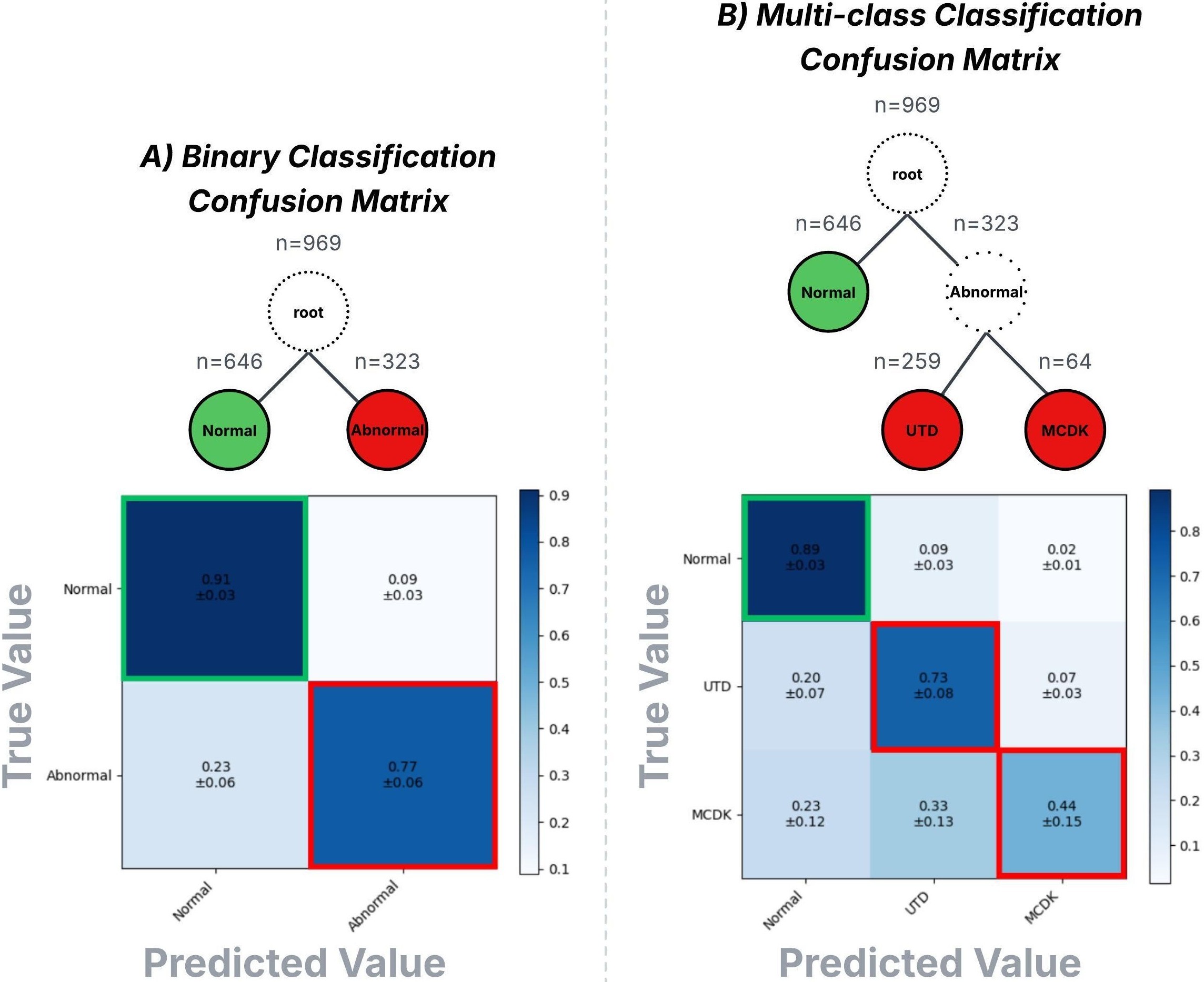}
	\caption{Confusion matrices for binary and multi-class renal anomaly classification.
    (A) Confusion matrix for the binary task, showing model performance in distinguishing normal from abnormal fetal kidneys.
    (B) Confusion matrix for the multi-class task, illustrating prediction distributions across normal, UTD, and MCDK categories. Values represent mean performance with variation across test folds.}
	\label{FIG:comparison_of_cm}
\end{figure*}

\section{Discussion}
In the present work, we demonstrated the feasibility of a deep learning framework for the detection of fetal renal anomalies on prenatal ultrasound images. In binary and multi-class classification settings, USF-MAE outperformed the DenseNet-169 baseline across several performance metrics, achieving a higher F1-score, greater discriminative power, and improved generalization. These results support the utility of ultrasound-specific self-supervised pretraining for enhanced downstream performance on obstetric imaging tasks.

\subsection{Improved Diagnostic Performance Across Tasks}
For the binary classification task, USF-MAE outperformed DenseNet-169 in almost all evaluated performance metrics. This held true on both the validation ({\hypersetup{hidelinks}\textcolor{blue}{Table~\ref{tab:val_set_binary}}}) and independent holdout test sets ({\hypersetup{hidelinks}\textcolor{blue}{Table~\ref{tab:test_set_binary}}}). Improvements were consistent across AUC, accuracy, F1-score, and precision, suggesting that the model provided a more robust separation between normal and abnormal fetal kidneys. Supporting this, the ROC and precision-recall curves in {\hypersetup{hidelinks}\textcolor{blue}{Figs.~\ref{FIG:performance_curves}A}} and {\hypersetup{hidelinks}\textcolor{blue}{\ref{FIG:performance_curves}B}} show high AUC and strong agreement between cross-validation folds.

We observed greater performance gains for the more challenging multi-class classification task. In this setting, model predictions included the additional classes of UTD and MCDK, and the USF-MAE framework was able to more clearly distinguish between these and normal cases. As shown in {\hypersetup{hidelinks}\textcolor{blue}{Table~\ref{tab:multi_class_combined}}}, this was reflected by substantial improvements in both accuracy and F1-score compared to DenseNet-169. These gains were particularly pronounced for the minority classes, where CNNs are known to show limitations. This can be clearly seen in the ROC and precision-recall curves in {\hypersetup{hidelinks}\textcolor{blue}{Figs.~\ref{FIG:performance_curves}C}} and {\hypersetup{hidelinks}\textcolor{blue}{\ref{FIG:performance_curves}D}}, respectively, which reveal that USF-MAE was able to distinguish between the three types of renal conditions and achieved good per-class ROC-AUC for both UTD and MCDK.

\begin{figure*}
	\centering
	\includegraphics[width=.98\textwidth]{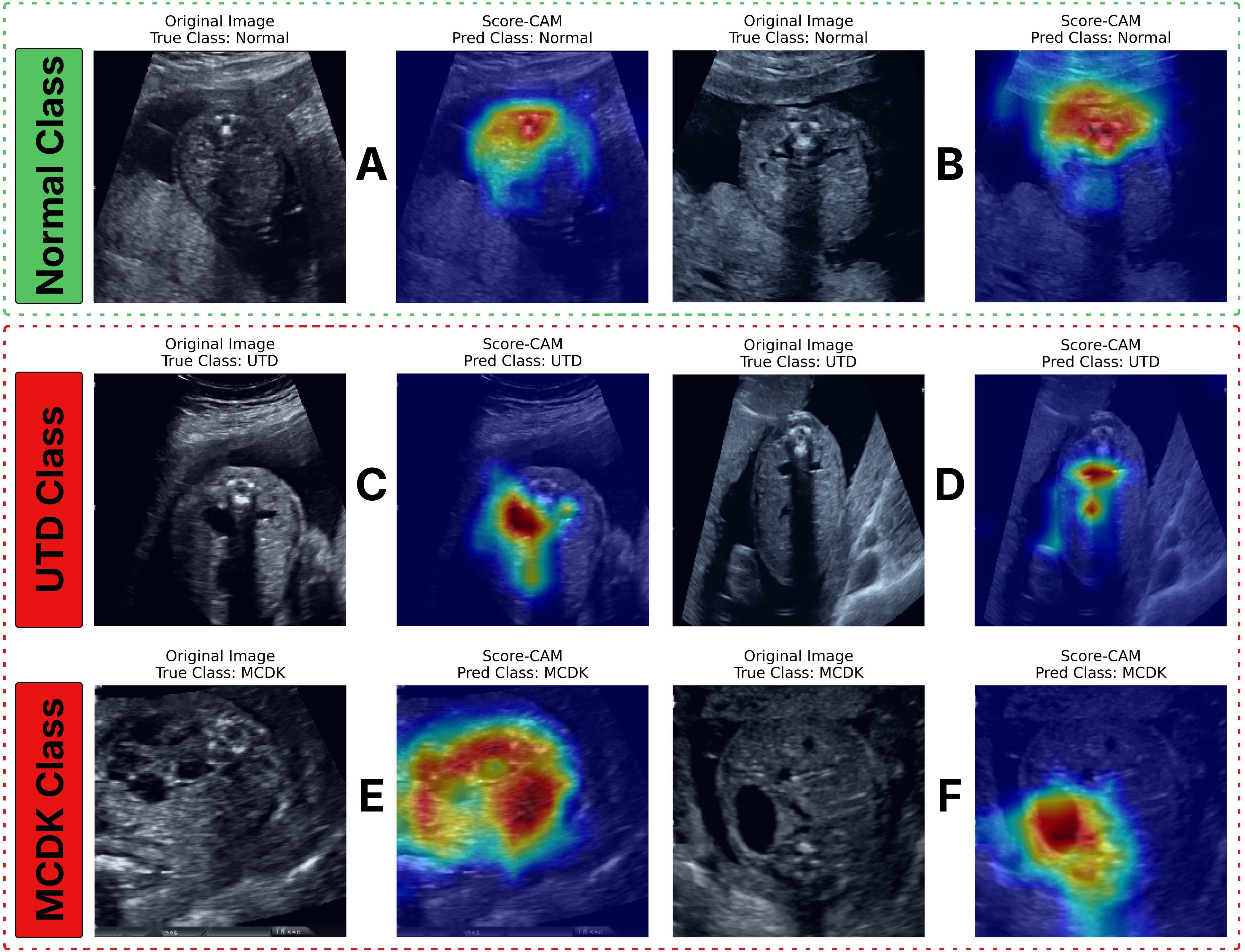}
	\caption{Score-CAM visualizations for model predictions across all classes. Examples are shown for normal, UTD, and MCDK cases, with each pair containing the original ultrasound image and its corresponding Score-CAM heatmap.
    (A-B) Normal class examples in which the model highlights regions consistent with typical renal anatomy.
    (C-D) UTD class examples, with attention concentrated around dilated collecting system features.
    (E-F) MCDK class examples, where the model focuses on the characteristic cystic structures. 
    These maps illustrate which image regions most strongly influenced the model’s predictions.}
	\label{FIG:XAI_figure}
\end{figure*}

\subsection{Interpretation Through Confusion Matrices}
Additional information about model performance is provided by the confusion matrices. In the binary case ({\hypersetup{hidelinks}\textcolor{blue}{Fig.~\ref{FIG:comparison_of_cm}A}}), USF-MAE exhibited high true negative and true positive rates, with mean correct classification values of $0.91 \pm 0.03$ for normal cases and $0.77 \pm 0.06$ for abnormal cases. This reflects a balanced detection profile that mitigated both false negatives and false positives.

In the multi-class task ({\hypersetup{hidelinks}\textcolor{blue}{Fig.~\ref{FIG:comparison_of_cm}B}}), the model also maintained high performance on the normal class and showed meaningful separation of UTD and MCDK. Classification of MCDK was more variable due to the limited sample size for this anomaly. Nonetheless, the USF-MAE still produced a correct classification rate of $0.44 \pm 0.15$ for MCDK and $0.73 \pm 0.08$ for UTD, suggesting that the model has learned clinically relevant patterns that can distinguish between different subtypes of anomalies.

\subsection{Explainability and Clinical Relevance}
Score-CAM {\hypersetup{hidelinks}\textcolor{blue}{\cite{b20}}} visualization of model attention for representative images from each class ({\hypersetup{hidelinks}\textcolor{blue}{Fig.~\ref{FIG:XAI_figure}}}) further indicates that model predictions were grounded in clinically meaningful patterns. In all three cases, Score-CAMs highlighted anatomically meaningful regions within the fetal kidneys as the area of highest attention. In the normal case, this overlapped with the expected renal outline and pelvicalyceal anatomy. In the UTD case, the CAM correctly focused on the dilated renal pelvis, capturing the defining feature of this anomaly. In the MCDK case, attention was concentrated on the multicystic kidney. These results emphasize that model predictions were not driven by spurious image regions or patterns.

\subsection{Overall Gains and Implications}
USF-MAE demonstrated a performance improvment of \textbf{1.87\%$\uparrow$} in AUC and \textbf{7.8\%$\uparrow$} in F1-score on the validation set, \textbf{2.32\%$\uparrow$} (AUC) and \textbf{4.33\%$\uparrow$} (F1-score) on the holdout test set, and \textbf{16.28\%$\uparrow$} (AUC) and \textbf{46.15\%$\uparrow$} (F1-score) in the multi-class setting. This is consistent with the hypothesis that the ultrasound-specific self-supervised pretraining step was beneficial for model performance. These improvements are also suggestive of meaningful added value that could be leveraged in clinical practice. In particular, the model showed strong performance in the more challenging multi-class setting, which could be leveraged to support early detection of fetal renal anomalies and identification of cases for follow-up imaging or specialist referral.

\subsection{Limitations}
Several limitations should be considered when interpreting the results of this study. First, although the dataset was collected across multiple tertiary care sites, all ultrasound images were acquired using GE Voluson systems. Differences in image appearance, beamforming, postprocessing, and noise characteristics across ultrasound vendors may affect model performance when applied to images from other manufacturers. While the self-supervised pretraining of USF-MAE leveraged a large and heterogeneous ultrasound corpus, external validation on data acquired from non-GE systems will be necessary to fully assess cross-vendor generalizability.

Second, the number of MCDK cases was relatively small compared to Normal and UTD cases. While class weighting and self-supervised pretraining mitigated some effects of class imbalance, the limited sample size may contribute to variability in class-specific performance metrics. Expanding the dataset to include more rare renal anomalies would further improve robustness.

\subsection{Future Work}
Future work should be directed towards examining the generalizability of this framework to other ultrasound datasets and ultrasound machines, ideally at scale using multi-institutional data from diverse clinical sites. Including images from different ultrasound machines or exam protocols would allow for more robust model evaluation. Extending the model to use temporal cine-loop data, or three-dimensional ultrasound volumes, may also help capture subtle anatomical features. Finally, incorporating uncertainty quantification into the model could improve its clinical applicability by flagging low-confidence cases that may warrant manual review.


\section{Conclusion}
In this work, we proposed a self-supervised ultrasound foundation model for fetal renal anomaly detection on prenatal ultrasound. Fine-tuning a pretrained USF-MAE encoder on our fetal renal dataset led to consistent and significant performance improvements over a convolutional baseline in both the binary and multi-class settings. Performance was evaluated using the standard AUC, accuracy, F1-score, precision, recall, and specificity metrics, with improvements across the board. These findings underscore the utility of ultrasound-specific self-supervised representation learning for subsequent fine-tuning in diagnostic tasks.

USF-MAE exhibited strong discriminative performance in both binary and multi-class classifications. In the binary task, the model achieved high performance in distinguishing between normal and abnormal kidneys. In the multi-class setting, USF-MAE was able to effectively differentiate between the two major subtypes of renal anomaly: UTD and MCDK. The corresponding ROC and precision-recall curves indicated a stable performance across cross-validation folds, while confusion matrices demonstrated a balanced detection of both normal and abnormal classes on the independent holdout test set. Furthermore, the use of Score-CAM visualizations suggested that the predictions made by the model were attributable to biologically meaningful regions of the kidney, thereby supporting the interpretability of the model.

Cumulatively across all reported metrics, USF-MAE exhibited performance improvements of about 7.8\%$\uparrow$ and 4.33\%$\uparrow$ in F1-score on the validation and independent holdout test sets, respectively, and a 46.15\%$\uparrow$ improvement in F1-score in the multi-class setting over the DenseNet-169 baseline. These results demonstrate that self-supervised pretraining on a large ultrasound corpus can meaningfully improve downstream model performance, especially in cases that require subtle anatomical distinctions and address data imbalance. Our research shows that self-supervised ultrasound foundation models can effectively detect prenatal anomalies and enhance early detection of fetal renal abnormalities while minimizing false positive detections. They may also help to direct more efficient and timely referrals for follow-up with a specialist.

\section*{Ethical Approval}
This study was reviewed and approved by the Ottawa Health Sciences Network Research Ethics Board (OHSN-REB \#20210079). All methods were performed in accordance with the relevant institutional guidelines and regulations and in alignment with the Tri-Council Policy Statement: Ethical Conduct for Research Involving Humans (TCPS-2). This study involved images previously collected at the study centre, which were de-identified before model training and validation. Seeking participant consent was waived by the Ottawa Health Sciences Network Research Ethics Board, as this retrospective study relied exclusively on secondary use of non-identifiable information. The data management and analysis for this study were conducted within the secure institutional network environment.

\section*{Availability of Data and Materials}
The original image files are not publicly available because they contain information that could compromise the privacy of research participants. The datasets generated and analyzed for this study are available from the corresponding author (MCW), upon reasonable request.

\section*{Funding}
This study was supported by a Canadian Institutes of Health Research Foundation Grant (FDN 148438). The funding agency was not involved in the study design, analysis, or interpretation of the data. The agency was not involved in the writing of this manuscript or in the decision to submit the article for publication.

\section*{Declaration of Competing Interest}
The authors declare that they have no known competing financial interests or personal relationships that could have appeared to influence the work reported in this paper.


\end{document}